\documentclass[runningheads]{llncs}
\usepackage[T1]{fontenc}
\usepackage{graphicx}
\usepackage{amsmath,amssymb,amsfonts}
\usepackage{subcaption}
\usepackage[hidelinks]{hyperref}
\usepackage[frozencache,cachedir=.]{minted}
\begin{document}

\captionsetup{font=small}
\captionsetup[sub]{font=small}
\title{EfiMon: A Process Analyser for Granular Power Consumption Prediction}
%

\author{Luis G. Le\'on-Vega\inst{1}\orcidID{0000-0002-3263-7853} \and Niccol\`o Tosato\inst{1,2}\orcidID{0009-0005-4996-9933} \and Stefano Cozzini\inst{2}\orcidID{0000-0001-6049-5242}}
\authorrunning{L. Le\'on-Vega et al.}
%
\institute{Università degli Studi di Trieste, Trieste, Italy \email{\{luisgerardo.leonvega,niccolo.tosato\}@phd.units.it} \\ \and AREA Science Park, Trieste, Italy \\
 \email{stefano.cozzini@areasciencepark.it}}

\maketitle              
\begin{abstract}
High-performance computing (HPC) and supercomputing are critical in Artificial Intelligence (AI) research, development, and deployment. The extensive use of supercomputers for training complex AI models, which can take from days to months, raises significant concerns about energy consumption and carbon emissions. Traditional methods for estimating the energy consumption of HPC workloads rely on metering reports from computing nodes’ power supply units, assuming exclusive use of the entire node. This assumption is increasingly untenable with the advent of next-generation supercomputers that share resources to accelerate workloads, as seen in initiatives like Acceleration as a Service (XaaS) and cloud computing.

This paper introduces EfiMon, an agnostic and non-invasive tool designed to extract detailed information about process execution, including instructions executed within specific time windows and CPU and RAM usage. Additionally, it captures comprehensive system metrics, such as power consumption reported by CPU sockets and PSUs. This data enables the development of prediction models to estimate the energy consumption of individual processes without requiring isolation.

Using a regression-based mathematical model, our tool is able to estimate single processes' power consumption in isolated and shared resource environments. In shared scenarios, the model demonstrates robust performance, deviating by a maximum of 2.2\% on Intel-based machines and 4.4\% on AMD systems compared to non-shared cases. This significant accuracy showcases EfiMon's potential for enhancing energy accounting in supercomputing, contributing to more efficient and energy-aware optimisation strategies in HPC.

\end{abstract}
\section{Introduction}

Artificial Intelligence (AI) is becoming the most demanding HPC workload as the models become more complex and compute-hungry, inspiring a transition of HPC to Acceleration as a Service (XaaS) using microservices~\cite{hoefler2024xaas}. Moreover, there is an increasing concern about the energy consumption utilised for AI in training and inference, given the large number of devices used for Large-Language Models (LLM). AI is also triggering large-scale investments in supercomputing, increasing power consumption exponentially to meet the demand, posing a risk in reducing carbon emissions. The Top 5 most powerful supercomputers require 7 MW up to 30 MW to reach the maximum capacity~\cite{top500}. In the case of Italy during 2022, 63.4\% of the electricity was produced by fuels and 27.5\% by renewable energies like wind, solar and hydro~\cite{iea}, posing an issue of how energy is produced to supply electricity in an increasing demand from HPC.

Currently, the accounting for energy in HPC workloads is limited to the node level, without a clear distinction of the actual computation, communication, and energy consumption distribution among the running processes. This lack of granularity hampers our understanding of how power-hungry the workloads are and which parts are the most energy-demanding. Our research aims to fill this knowledge gap, leading to more efficient and energy-aware optimisation strategies.

This work aims to contribute with a tool, called \emph{EfiMon}, capable of gathering process information about its load footprint and overall system information about its overall occupancy and power consumption. By gathering this information, it is possible to propose new prediction models that can estimate the energy spent by a Process of Interest (PoI) without the need for execution isolation, making them ideal for shared computing measurements. Additionally, we tested our tool by fitting a regression model proposed in previous work to show the effectiveness of the information compiled by EfiMon for non-isolated granular energy estimation.

\section{Background and Related Work}

Currently, the computing vendors roughly equip the server hardware with tools to explore and analyse the energy consumption in HPC facilities. This results in limitations in the instruments available in the computing nodes when quantifying the energy expense and restricts the possibility of deciding when and how to allocate resources to execute a particular job. For instance, Intel's tools, such as PCM and RAPL, provide measurements of aggregate core power consumption, socket consumption as an aggregation of all cores plus the non-processing components, and DRAM consumption~\cite{psti}. In contrast, AMD has a more limited set of tools, excluding the DRAM from the measurements~\cite{psti}, and some tools like AMD uProf are closed-source~\cite{uprof}, restricting the analysis of energy consumption at the workload level on these systems.

SLURM~\cite{slurm}, the most popular job scheduling software, does not optimise job scheduling based on energy consumption or make any decisions based on an energy budget. Instead, it only provides the energy consumed system-wide by a job based on the energy meters available, like IPMI and Intel RAPL, and turns on or off the nodes to save energy.

One of the main concerns regarding SLURM's approach is that it fails to account for the energy consumption of individual processes or tasks unless they are running in complete isolation within the system. This is because Intel RAPL and IPMI operate at a system-wide level. In some cases, these tools may also be inaccessible or inaccurate, leaving a gap in the available metrics. To address this, other methods have been proposed to approximate the power consumption for server computers, such as CPU utilisation, temperature, and fan speed as input for estimations~\cite{power-fanspeed,energy-temp,sthocastic-model,power-base-mem-access,power-usage-var}. Some researchers have also suggested more accurate measurement approaches by using stochastic methods or probabilistic models to describe the system as a whole~\cite{sthocastic-model,energy-law-issue}, which have shown promising results in experiments.

Contributions that aim to analyse energy consumption at a fine-grained level typically focus on Embedded Systems that use ARM microprocessors or RISC microcontrollers, as they are often used in battery-powered applications. These studies have examined the impact of executing certain instructions, such as load/storage in ARM~\cite{arm-load-store}, Very Long Instruction Words (VLIW) and memory address segments~\cite{vliw-immediate-inst}. Some proposals include power models to extract energy consumption based on the executed instructions, while others suggest using energy heuristics during development and compile time~\cite{dev-compile}.

When switching to server-level computers, the scenario abruptly changes with increased complexity, and some fine-grain energy strategies are insufficient and require more study. An integral energy analysis includes cooling, storage, networking, memory and multi-socket CPU consumption. A survey from 2016 suggests that 50\% is invested in cooling, 10\% in storage and 10\% in networking~\cite{survey-2016}. On the other hand, a survey from 2020 reported that the CPU takes 32\% of the power consumption~\cite{survey-2020}. Despite the changes in the data, energy accounting implies more components than analysing the CPU power consumption.

The community has addressed estimating the energy consumed by server computers mainly using models with parameters such as CPU utilisation and frequency, memory and disk utilisation, temperature, fan speed, and throughput (bandwidths)~\cite{survey-2020}. Our key finding during our state-of-the-art study is that most models try to estimate the energy at a system-wide level, and there has been little recent work on models that use instructions executed by workloads as parameters, as found in embedded systems. Therefore, there is an opportunity to contribute to the field of process-level energy consumption (fine-grained) using the executed instructions as a parameter for the model to enhance the energy consumption estimation~\cite{efimon-model}.

This work will focus on proposing a tool to gather information to fit a regression model that takes into account the instructions executed by a PoI, and its load on the system. This will also analyse the effectiveness of the information provided by the tool in making predictions in isolated and non-isolated execution environments.

\section{Instruction-Based Power Analysis}

Our contribution consists of two main sections: 1) an agnostic performance monitoring tool for measuring the system's energy consumption, CPU utilisation, and instructions executed by a process, and 2) the analysis of a model to measure the power consumption of a PoI based on the instructions executed and the CPU utilisation.

\subsection{Efimon: Agnostic Performance Monitor}

Our work proposes a tool composed of a C++ library and applications to extract information from the process execution and the overall system metrics. From the process perspective, EfiMon extracts the instructions executed, sampled at a certain rate during a time window, and includes the CPU, RAM and network utilisation. From the system perspective, it extracts critical information about power consumption from the CPU sockets and PSU, as well as the overall system load with respect to the CPU, RAM, GPU and network.

EfiMon's library follows the interface-adapter architecture combined with the observer pattern. The architecture allows EfiMon to standardise the API, making it extensible through the adapter implementation. For instance, CPU socket power consumption can be obtained from Intel PCM, RAPL or AMD uProf. Depending on the system where it is compiled, the adapters are enabled, offering the possibility to get measurements from any of them while preserving the API, making negligible changes to the final application. 
\begin{figure}[t]
    \centering
    \includegraphics[width=0.95\columnwidth]{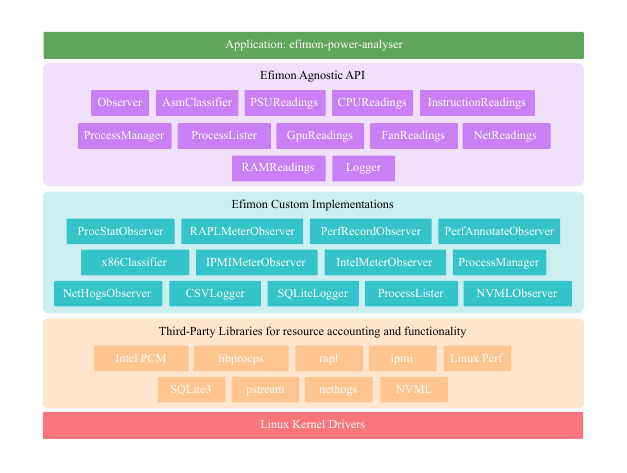}
    \vspace{-5mm}
    \caption{Efimon software stack. The architecture uses the interface-adapter pattern to decouple the dependencies and keep the API uniform and agnostic.}
    \vspace{-5mm}
    \label{fig:sw-stack}
\end{figure}

Fig.~\ref{fig:sw-stack} shows the software architecture of the EfiMon project, summarising the different levels of abstractions from the hardware drivers up to the final user application. From bottom to top, EfiMon's software stack uses Linux libraries like \emph{libprocps} to access the process statistics \emph{Linux Perf} to access kernel events and hardware counters, the \emph{RAPL} interface and \emph{Intel PCM} to get information about the CPU energy statistics and \emph{IPMI} for the PSU mean power and fan speeds. For housekeeping and auxiliary functions, EfiMon uses \emph{SQLite3} to save recordings in a database file and \emph{pstream} to process deployment and signalling. EfiMon also includes tools to measure network traffic using \emph{NetHogs} and GPU statistics using \emph{NVML} for a specific process; however, they will be analysed in future work.

For abstractions, we propose two user-accessible interface classes called \emph{Observer} and \emph{Readings}. The \emph{Observer} standardises basic functionalities such as reading results, triggering metering and common constructors that set the process ID, sampling frequency, metering interval and reading scope. The Observer interface class is implemented using adapters, concrete classes that wrap up the dependencies mentioned above. Table~\ref{tbl:dependencies} shows all the implementations in EfiMon. Each implementation uses a single dependency, helping with conditional compilation and availability of tools after the project's construction. The \emph{Readings} specialisable structure is a base data container that sets basic members such as timestamp and time difference. This structure is intended to be extendible to include measurements, depending on the observer. Each \emph{Observer} adapter fills a specialisation of the Readings class, standardising the metrics' storage and keeping the API uniform across observers. For instance, the \emph{ProcStatObserver} and the \emph{RAPLMeterObserver} specialise the \emph{Readings} class into \emph{CPUReadings} to hold the CPU metrics. The \emph{Readings} subclasses can have unfilled members. In the case of \emph{ProcStatObserver}, it only fills the CPU utilisation members, whereas the \emph{RAPLMeterObserver} the power measurements.

\begin{table}[t]
\centering
\caption{Observer implementations and their readings specialisation}
\scriptsize
\begin{tabular}{|l|c|c|}
\hline
\textbf{Observer Implementation} & \textbf{Dependency Used} & \textbf{Readings Implemented} \\[3pt] \hline
ProcStatObserver                 & libprocps                & CPUReadings, RAMReadings      \\[3pt] \hline
RAPLMeterObserver                & rapl                     & CPUReadings                   \\[3pt] \hline
PerfRecordObserver               & perf                     & N/A                           \\[3pt] \hline
PerfAnnotateObserver             & perf, x86Classifier      & InstructionReadings           \\[3pt] \hline
IPMIMeterObserver                & ipmi                     & FanReadings, PSUReadings      \\[3pt] \hline
IntelMeterObserver               & Intel PCM                & CPUReadings                   \\[3pt] \hline
NetHogsObserver                  & nethogs                  & NetReadings                   \\[3pt] \hline
NVMLObserver                     & nvml                     & GPUReadings                   \\[3pt] \hline
\end{tabular}
\label{tbl:dependencies}
\end{table}

\begin{listing}[!ht]
\vspace{-5mm}
\caption{Oversimplified example of the EfiMon's library usage for RAPL, IPMI and libprocps measurements.}
\begin{minted}
[
frame=lines,
framesep=2mm,
baselinestretch=1.2,
fontsize=\scriptsize,
linenos
]
{c++}
// Create Observers
auto rapl_obs = RAPLMeterObserver{};
auto ipmi_obs = IPMIMeterObserver{};
auto proc_obs = ProcStatObserver{pid, efimon::ObserverScope::PROCESS};

// Trigger a reading
rapl_obs.Trigger();
ipmi_obs.Trigger();
proc_obs.Trigger();

// Get readings
CPUReadings *cpu_readings =
    dynamic_cast<CPUReadings *>(rapl_obs.GetReadings()[0]);
PSUReadings *psu_readings =
    dynamic_cast<PSUReadings *>(ipmi_obs.GetReadings()[0]);
FanReadings *fan_readings =
    dynamic_cast<FanReadings *>(ipmi_obs.GetReadings()[1]);
CPUReadings *cpu_readings_pid =
    dynamic_cast<CPUReadings *>(proc_obs.GetReadings()[0]);
RAMReadings *ram_readings_pid =
    dynamic_cast<RAMReadings *>(proc_obs.GetReadings()[1]);

// Accessing to a member
float psu_power = -1.f;
if (psu_readings) psu_power = psu_readings->psu_power.at(0);
\end{minted}
\label{algo:pseudocode}
\vspace{-6mm}
\end{listing}

Listing~\ref{algo:pseudocode} shows an oversimplified example of using EfiMon's API within a metering application. Lines 2 and 3 show how to create system-wide observers, and line 4 shows how to create a process-scoped observer (attached to \texttt{pid}). By default, if the constructor is left empty, it sets the PID and the interval to 0 and the scope to the default for the observer. This may vary from one observer to another.

The \texttt{Trigger} method performs a reading to the adapter, gathering the information about the CPU socket (line 7), IPMI (line 8) and libprocs (line 9). Internally, this method queries all the respective sensors or functions to refresh the internal measurements for later use with the \texttt{GetReadings} method.

The \texttt{GetReadings} method returns a vector of pointers to \emph{Readings}, whose memory segment points to the internal readings of the observer instance. To access the observer's measurements, the \emph{Readings} pointer type must be cast to the actual subclass, as illustrated in lines 12-21, containing the members for a specific measurement (line 25). Using this approach, the API is standardised, and the readings can include polymorphism, making them flexible for future extensions.

The tool is available online on Zenodo~\cite{efimon-sw}.

\subsection{Process Energy Model}

The instantaneous power consumption of a server computer, at an instant $t$, can be modelled as the sum of the power consumption of the CPU, hardware accelerators (i.e. GPUs), RAM, storage, network interface cards (NIC), cooling (fans) and other electronic components~\cite{power-addition}:
\begin{multline}
    P_{\text{system},t} = P_{\text{CPU,t}} + P_{\text{Accel,t}} + P_{\text{RAM,t}} + P_{\text{Disk,t}} + P_{\text{NIC,t}} + P_{\text{Cool,t}} + P_{\text{Aux,t}}
\end{multline}

We can also express the total system power as the power supplied by the Power Supply Unit (PSU), and the rest of the power components as a composition of static power and dynamic power, where the latter depends on the process computational activity~\cite{power-core-decomp}.
\begin{equation}
    P_{\text{system},t} = P_{\text{PSU},t} = P_{\text{static},t} + P_{\text{dynamic},t}
    \label{eqn:system-dyn-st-power}
\end{equation}

The static power comprises all the devices unaffected by the computing activity and remains fixated during the execution, like fans running at a constant speed and unused or suspended devices~\cite{power-base-mem-access}. For the scope of this work, the fan speed is fixed at its maximum speed.

The dynamic power component depends on computing activity, which includes contributions from all running processes. Each contribution depends on core frequency, instructions executed, degree of parallelism, and CPU, RAM, network and accelerator utilisation~\cite{efimon-model}.

Our previous work analyses the decomposition into processing-dependent and static power, implicitly covering CPU, RAM and other peripherals unrelated to computations but essential for execution, such as memory movements. We have also analysed the power model of the dynamic power component, assuming fixed CPU clock frequency and fan speed. Instructions are classified into types (scalar and vector) and families (arithmetic, logic, memory, branches and jumps), with each instruction type and family contributing non-linearly to dynamic power. To simplify the analysis, we compress the instruction type and family proportions in a histogram $h_k$, resulting in the following formulation:
\begin{equation}
    P_{\text{dynamic},t} = \sum_{p=1}^{N_p}\sum_{k=1}^{S} \gamma_{k} \sigma(h_k^{p}, w_p)
    \label{eqn:dynamic-model-process}
\end{equation}

\noindent
where $N_p$ is the number of running processes in the system, $S$ is the number of instruction types/families executed by the process $p$ in the time sample $t$, $\gamma_k$ is the weight of each instruction type on overall consumption, and $\sigma$ is a function that describes the relation between each instruction type/family proportion $h_{k}^{p}$ and the process utilisation $w_p$.

Merging equations (\ref{eqn:system-dyn-st-power}) and (\ref{eqn:dynamic-model-process}), the entire system power can be written as:
\begin{equation}
    P_{\text{system},t} = P_{\text{static},t} + \sum_{p=1}^{N_p}\sum_{k=1}^{S} \gamma_{k} \sigma(h_k^{p}, w_p)
    \label{eqn:system-model-process}
\end{equation}

For the regression model, we will use the $P_{\text{PSU},t}$ as the reference value and modify the model to accept a single process running in isolation. This is possible thanks to the conservation of energy and the superposition analysis, which dictates that the total supplied power is the sum of all the power loads independently:
\begin{equation}
    P_{\text{PSU},t} = \sum_{k=1}^{S} \gamma_{k} \sigma(h_k^{p}, w_p) + \hat{P}_{\text{static},t}
    \label{eqn:regression-model-process}
\end{equation}

\noindent
where $\hat{P}_{\text{static},t}$ is the intercept of a linear regression model approximated during the model fitting.

\section{Experimental Results}

We aim to observe the behaviour of CPU utilisation and the impact of the instruction types on the overall power consumption. Additionally, we propose a power model dependent on CPU consumption and instruction type probabilities for a given process. First, we use EfiMon to capture the system and process CPU utilisation, the server power consumption, and a histogram of instruction type probabilities. Then, we analyse the results obtained from EfiMon, discussing the behaviour of each independent variable on the overall CPU power consumption and providing insights into the nature of the function $\sigma$. Last, we will apply a non-negative linear regression to find the model's parameters.

For experimental data, we use two dedicated nodes from the ORFEO cluster configured as reported in Table~\ref{tbl:hardware}. To reduce the noise during the measures, the clock speeds have been fixed at 1.5 GHz (AMD) and 1.0 GHz (Intel), C-states and boost are disabled, and the fan speed is set to maximum. 
\begin{table}[t]
\scriptsize
\centering
\caption{Hardware Configuration}
\begin{tabular}{|l|c|c|}
\hline
\textbf{Comp.} & \textbf{AMD-based Node} & \textbf{Intel-based Node}                         \\[3pt] \hline
CPU & 2 X AMD EPYC 7H12 64 cores                        &  2 X Intel(R) Xeon(R) Gold 6126 CPU @ 2.60GHz \\[3pt] \hline
RAM     & 16 X 32GB 3200 MT/s DDR4 ECC                  &  	12 X 64GB 2666 MT/s DDR4 ECC  \\[3pt] \hline
Disk & 2 X 480GB SSD -  RAID 0                          &  2 X 480GB SSD -  RAID 0  \\[3pt] \hline
NIC               & MT28908 Family 100Gb (Infiniband)   & MT28908 Family 100Gb (Infiniband) \\[3pt] \hline
Cooling                & 16 x Dell (Silver grade) fan   & 16 x Dell (High Performance) fan \\[3pt] \hline
PSU                &  2 x 1400 W redundant PSU          &  2 x 1100 W redundant PSU \\[3pt] \hline
\end{tabular}
\label{tbl:hardware}
\end{table}
Likewise, the experiments performed are the following:

\begin{itemize}
    \item \textbf{idle}: measures the idle power of the system when executing nothing.
    \item \textbf{copy}: performs memory copies using scalar instructions.
    \item \textbf{copy\_mem} and \textbf{copy\_mem\_avx}: performs memory copies using scalar and vector units (AVX) but avoiding temporal storage, exploiting the bandwidth in NUMA machines.
    \item \textbf{daxpy\_mem\_avx\_fma}: a double-precision linear combination of two vectors optimised for AVX FMAs and non-temporal stores.
    \item \textbf{stream\_mem\_avx\_fma}: a double-precision stream triad $a_i = k b_i + c_i$, optimised for AVX FMAs and non-temporal stores.
    \item \textbf{load}: double-precision memory loads using non-temporal data and AVX vector instructions.
    \item \textbf{store}: double-precision memory stores using non-temporal data and AVX vector instructions.
    \item \textbf{peakflops\_avx\_fma}: double-precision multiplications and additions with a single load, optimised for AVX FMA.
    \item \textbf{update\_avx}: AVX double-precision vector update (load and store).
    \item \textbf{dgemm}: double-precision matrix multiplication from Level 3 BLAS.
\end{itemize}

These experiments imply multi-threading using OpenMP to stress power consumption during execution. All benchmarks can be found in the likwid-benchmark utility~\cite{psti}, and the \textbf{dgemm} is implemented with the OpenBLAS library~\cite{openBLAS}. Experiments using hardware acceleration and Inter-Process Communication (IPC) within the node and over the network are excluded from the scope of this work and will be covered in future contributions.

\subsection{Preliminary Experimental Study}

\begin{figure}[!h]
    \centering
    \includegraphics[width=0.85\columnwidth]{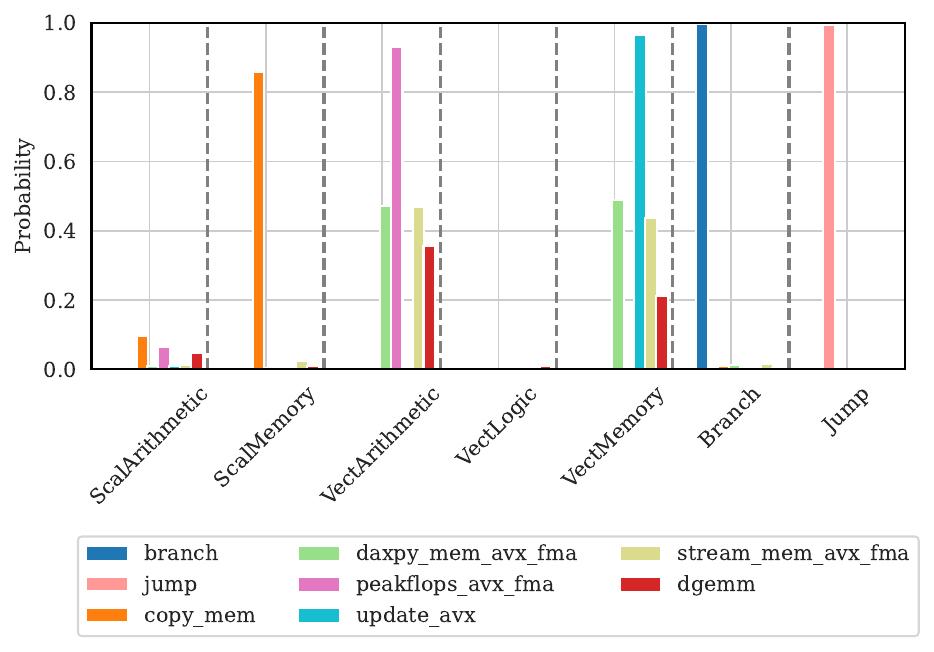}
    \caption{Mean probability of the instruction types for each experiment measured using Linux Perf through EfiMon.}
    \label{fig:historgram}
    \vspace{-5mm}
\end{figure}

The experiments present several scenarios given the instructions exercised during execution, implying a change in the probability distribution of instruction $\mathbf{h}$ (see Fig.~\ref{fig:historgram}). Additionally, multi-threading changes CPU utilisation $w$. Power consumption is measured from the socket using the RAPL interface.

\begin{figure}[!h]
    \begin{subfigure}[b]{0.48\columnwidth}
        \centering
        \includegraphics[width=0.95\columnwidth]{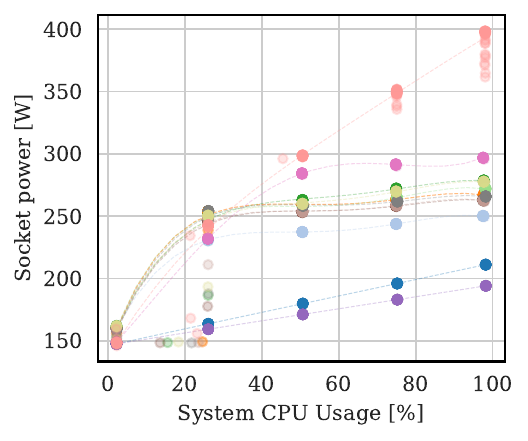}
        \caption{Total CPU Power}
        \label{fig:power-consumption-system-usage}
    \end{subfigure}\hfill
    \begin{subfigure}[b]{0.48\columnwidth}
        \centering
        \includegraphics[width=0.95\columnwidth]{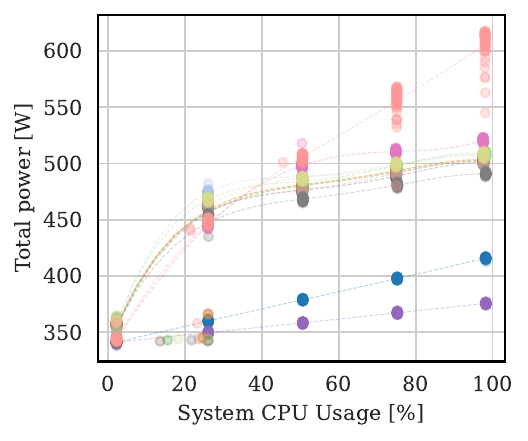}
        \caption{Total PSU Power}
        \label{fig:psu-consumption-system-usage}
    \end{subfigure}\\
    \begin{subfigure}[b]{\columnwidth}
        \centering
        \includegraphics[width=0.95\columnwidth]{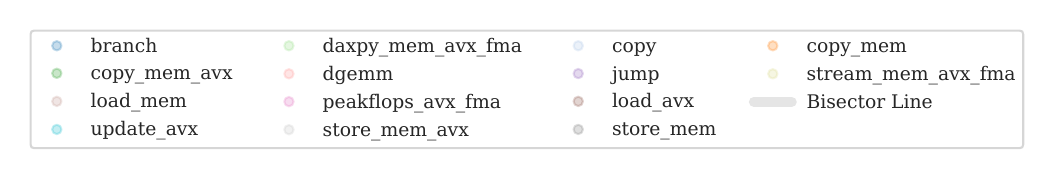}
    \end{subfigure}
    \caption{Total CPU and PSU consumption of all experiments at different CPU usage from 0 to 100\% (multiple threads). The total power comprises the two CPU sockets.}
    \vspace{-5mm}
\end{figure}

Fig.~\ref{fig:power-consumption-system-usage} and~\ref{fig:psu-consumption-system-usage} illustrate how CPU usage increases overall socket and PSU power consumption. The CPU usage is adjusted by adding more threads to the benchmark. Taking the system usage as an independent variable of the model, it has a logarithmic effect on the overall CPU and PSU power consumption. Each experiment also shows different power consumption at the same system CPU usage, suggesting that the instruction type also plays a role in the power consumption estimation. The plot helps to determine a non-linear relationship between experiment variables and power consumption, as experimental points scatter more with increasing system CPU usage.

According to the observations and our previous work~\cite{efimon-model}, we propose that $\sigma$ can be modelled for the memory transactions the vector arithmetic as:
\begin{equation}
    \sigma(h_k, w_p) = h_k^p\ln(N_cw_p + 1)
    \label{eqn:sigma-def}
a[a\end{equation}

\noindent
and for the rest of the instruction types as:
\begin{equation}
    \sigma(h_k, w_p) = N_ch_k^pw_p
    \label{eqn:sigma-def-2}
\end{equation}

\noindent
with $N_c$ as the number of cores, $h_k^p$ as the probability of the instruction type $k$ of a process $p$ and $w_p$ as the total CPU usage between 0 to 1 of the process $p$.

\subsection{Model Proposal and Analysis}

With $\sigma$ defined, we perform a linear regression using the equation (\ref{eqn:regression-model-process}) to find $\hat{P}_{\text{static},t}$ and the parameters $\gamma_k$. We fit the linear regression estimator with the total PSU power consumption (adding both PSUs), the histogram, process CPU utilisation, and the number of cores.

\begin{table}[t]
\centering
\caption{Parameters Estimated using NNLS Regression for CPU models using PSU Power Model from equation (\ref{eqn:regression-model-process}) as prediction reference}
\vspace{3mm}
\scriptsize
\begin{tabular}{|l|c|c|}
\hline
\textbf{Parameter Name}                     & \textbf{Power Model (AMD)}   & \textbf{Power Model (Intel)}    \\[3pt] \hline
Intercept ($\hat{P}_{\text{static},t})$     & 336.5031                   & 282.934     \\[3pt] \hline
Weight of Scalar Arithmetic ($\gamma_{sa}$) & 0.6717                     & 0.000    \\[3pt] \hline
Weight of Scalar Memory ($\gamma_{sm}$)     & 35.6589                    & 41.117     \\[3pt] \hline
Weight of Scalar Logic ($\gamma_{sl}$)      & 0.00000                    & 1.3363     \\[3pt] \hline
Weight of Vector Arithmetic ($\gamma_{va}$) & 38.6822                    & 34.510      \\[3pt] \hline
Weight of Vector Memory ($\gamma_{vm}$)     & 35.3435                    & 42.242    \\[3pt] \hline
Weight of Vector Logic ($\gamma_{vl}$)      & 154.5258                   & 59.149      \\[3pt] \hline
Weight of Branch ($\gamma_{vl}$)            & 0.6459                  & 10.800     \\[3pt] \hline
Weight of Jumps ($\gamma_{vl}$)             & 0.3239                  & 0.00000     \\[3pt] \hline
\end{tabular}
\label{tbl:parameters}
\end{table}

\begin{figure}[!h]
    \centering
    \begin{subfigure}[b]{0.49\textwidth}
        \centering
        \includegraphics[width=0.95\columnwidth]{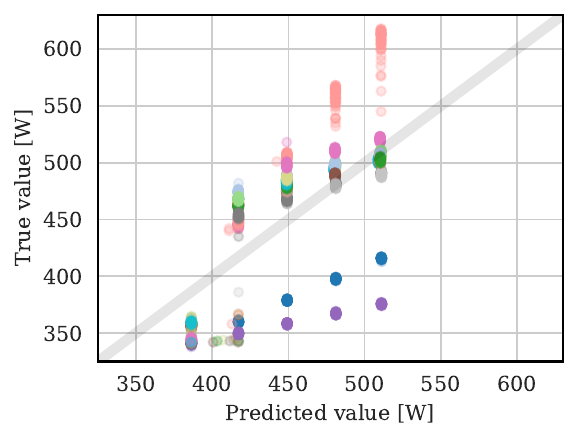}
        \caption{Usage only}
        \label{subfig:model-cpu-usage}
    \end{subfigure}\hfill
    \begin{subfigure}[b]{0.49\textwidth}
        \centering
        \includegraphics[width=0.95\columnwidth]{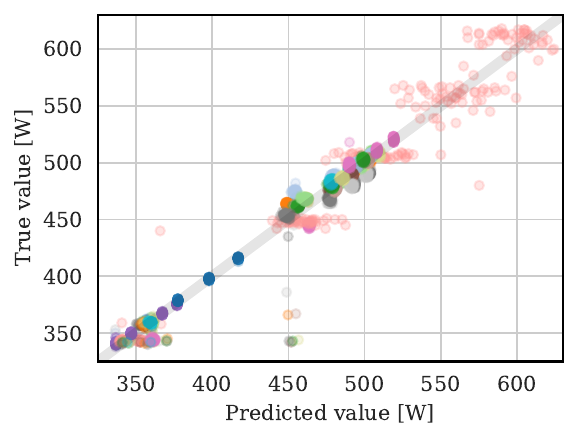}
        \caption{Usage + Instructions}
        \label{subfig:model-psu}
    \end{subfigure}\\
    \begin{subfigure}[b]{0.95\textwidth}
        \centering
        \includegraphics[width=\columnwidth]{figs/legend.pdf}
    \end{subfigure}\\
    \caption{Model prediction performances of different power consumption estimation models contrasted against the true value from IPMI (PSU) measurements. All models consider the CPU usage logarithmic.}
    \vspace{-5mm}
\end{figure}

Table~\ref{tbl:parameters} shows the non-negative least square regression (NNLS) results for eight types of instructions using the PSU as the power consumption reference and the AMD and Intel-based nodes. Other types were negligible and close to zero. Two outstanding weights were given to the vector operations followed by the memory transactions. This is expected since the CPU vector consumes most of the energy~\cite{efimon-model}, followed by the memory, where the memory transactions involve interacting with the DMA controllers and DRAM, causing significant dynamic power drain. This also matches with findings from other authors~\cite{arm-load-store} who have studied the effect of memory movements. The intercept compresses the entire system's static power, including fans, control hardware, and unused hardware such as the NIC and secondary storage. On the Intel architecture, the $\sigma$ changes the behaviour on the instructions, having a more linear behaviour similar to $\sigma_2$. When setting the $\sigma$ function for all instruction types, the error reduces from 14 Watts (with 3.2\% of central relative error) to 11 Watts (2.8\% error) and improves the predictions. On the AMD architecture, the $\sigma$ is preserved as explained in~\cite{efimon-model}, getting 9 Watts of error (less than 2\% error). It results in a hardware dependency of the $\sigma$ function on the CPU micro-architecture.

Fig.~\ref{subfig:model-cpu-usage} and~\ref{subfig:model-psu} illustrate the models using only the natural logarithm of the CPU usage as a parameter (setting $h_0^p = 1, h_k^p = 0; k \ne 0$) and the proposed model using the parameters from Table~\ref{tbl:parameters}. The model with CPU usage variance, as seen in Fig.~\ref{fig:power-consumption-system-usage}, is illustrated through same-color vertical point alignment, indicating the lack of parameters within the model. Fig.~\ref{subfig:model-psu} illustrates how the proposed model performs in terms of prediction. It demonstrates that the PSU model outperforms the former model, removing the variance among experiments and suggesting a better fit. The model is robust in most experiments but tends to have more variance when performing the \textbf{dgemm} experiment, which involves both memory and arithmetic, suggesting the presence of disturbations in the measurements performed by the tool and affecting the model prediction.

\subsection{Multi-Process Analysis}

One novel contribution of our work is quantifying energy consumed by a single process without the need for isolation. This section presents how our model and tool perform when measuring three different workloads of interest (PoI) while running other unrelated processes on the computing nodes specified in Table~\ref{tbl:hardware}. We maintain the same experiments conditions, fixing the frequency and the fan speed. The PoIs execute the \textbf{peakflops\_avx\_fma} (\textbf{peakflops}),  \textbf{daxpy\_mem\_avx\_fma} (\textbf{daxpy}) and \textbf{stream\_mem\_avx\_fma} (\textbf{stream}) with different numbers of cores and fixed workload size. We include a \textbf{dgemm} computation as a noise processing occupying 1/4\textsuperscript{th} of the total machine occupation. Regarding the model and parameters, we use the model from equation (\ref{eqn:regression-model-process}) with the parameters PSU-based NNLS of Table~\ref{tbl:parameters} without the intercept $\hat{P}_{\text{other},t}$, given their performance and generalisation over all the system power consumption.

\begin{figure}[!h]
    \centering
    \begin{subfigure}[b]{0.34\textwidth}
        \includegraphics[width=\columnwidth]{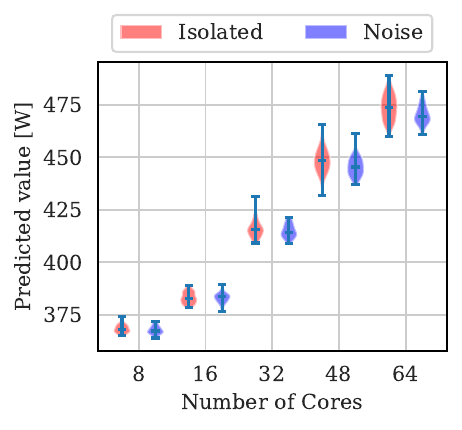}
        \caption{AMD - \textbf{stream}}
        \label{fig:experiment-comparison}
    \end{subfigure}\hfill
    \begin{subfigure}[b]{0.32\textwidth}
        \includegraphics[width=\columnwidth]{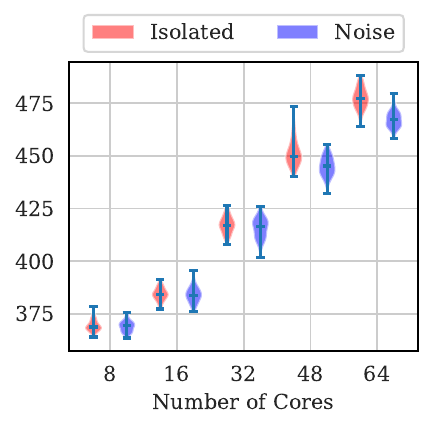}
        \caption{AMD - \textbf{daxpy}}
        \label{fig:experiment-comparison}
    \end{subfigure}\hfill
    \begin{subfigure}[b]{0.32\textwidth}
        \includegraphics[width=\columnwidth]{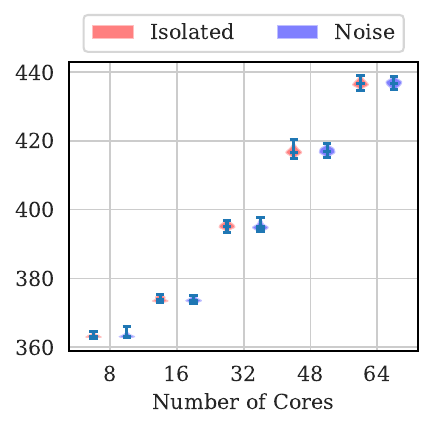}
        \caption{AMD - \textbf{peakflops}}
        \label{fig:experiment-comparison}
    \end{subfigure}\\
    \begin{subfigure}[b]{0.34\textwidth}
        \includegraphics[width=\columnwidth]{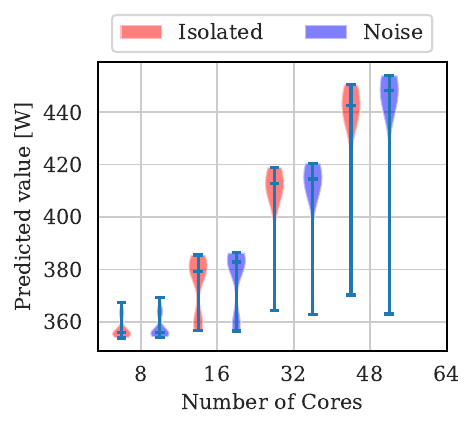}
        \caption{Intel - \textbf{stream}}
        \label{fig:experiment-comparison}
    \end{subfigure}\hfill
    \begin{subfigure}[b]{0.32\textwidth}
        \includegraphics[width=\columnwidth]{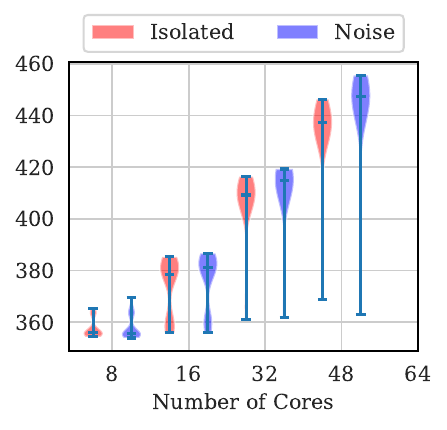}
        \caption{Intel - \textbf{daxpy}}
        \label{fig:experiment-comparison}
    \end{subfigure}\hfill
    \begin{subfigure}[b]{0.32\textwidth}
        \includegraphics[width=\columnwidth]{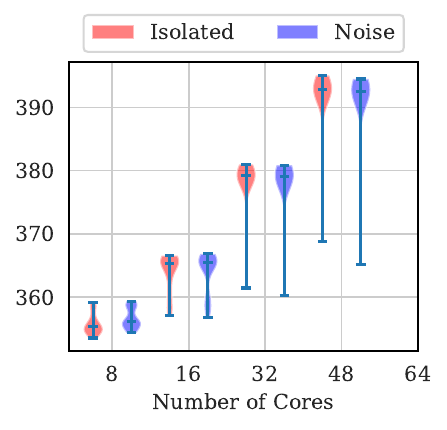}
        \caption{Intel - \textbf{peakflops}}
        \label{fig:experiment-comparison}
    \end{subfigure}
    \caption{Assessment of the prediction in the absence and presence of noise for different levels of parallelism (number of cores).}
    \vspace{-7mm}
    \label{fig:non-isolated-measurements}
\end{figure}

Fig.~\ref{fig:non-isolated-measurements} shows an assessment of the tool and the model when predicting the power consumption of workloads (organised in columns) on two different compute machines equipped with AMD and Intel. The figures illustrate two execution cases: 1) the PoI running in isolation and 2) the workload running with the \textbf{dgemm} to break the isolation (noise), all while scaling the PoI in the number of cores. The medians for a fixed number of cores and within the same PoI are close, giving the same power prediction for isolated and noise cases. In the AMD-based node, the \textbf{daxpy} has the maximum deviations when executing with 48 cores and 64 cores, with 15.4 and 20.7 Watts, corresponding to a relative error of 4.6\% and 4.4\% with respect to the isolated measurement, respectively. The best scores were achieved by the \textbf{peakflops} workload with a maximum deviation of 0.6 Watts. The key difference among the experiments relies on memory access. \textbf{daxpy} and \textbf{stream} have more memory accesses than the \textbf{peakflops}, which is a pure computational workload.

In the Intel-based node, the distributions are larger, but the behaviour is preserved within the experiment pairs, which implies that the behaviour of the noise environment comes from the measurements and not from the model. The medians preserve the alignment, suggesting that the predictions from the noise experiments are close to the isolation counterpart. Likewise, \textbf{daxpy} remains the most deviated as the number of cores increases, with a maximum deviation of 9.67 Watts (2.2\% of relative error) when running at 64 cores.

The results show that the model and tool are robust against noise, keeping a low gap between the noise and no-noise measurements, and the law of preservation of energy (from equation (\ref{eqn:regression-model-process})) is achieved with our proposed model and parameters. The experiments also suggest that our tool is promising in fine-grained energy accounting, contributing to the future of the HPC, where the directions point to sharing computational resources. There are still variables to study, such as frequency and fan speed, to make our tool more robust and complete.

\section{Conclusions and Future Work}

This study presents EfiMon, a robust tool for instruction-driven process energy consumption analysis in multi-socket computers within shared computational environments. Our contributions include an analysis tool capable of acquiring detailed power measurements from the CPU and PSU, a histogram of the instructions executed by a process of interest, and process usage metrics. The tool is built on top of our open-source library, leveraging an interface-adapter architecture based on observers to decouple dependencies and maintain a uniform API for usability.

Our work demonstrated robustness in predicting the power consumption of individual processes, even in shared resource environments. The regression model, trained on two different dual-socket compute nodes using a set of likwid benchmarks in isolated and shared scenarios, exhibited minimal deviation, 4.4\% for AMD and 2.2\% for Intel, validating its accuracy and reliability in predicting process power consumption.

Our results show EfiMon's promising potential in fine-grained energy consumption measurement, particularly in scenarios where computational resources are shared. This robustness and quality of prediction in energy estimation contribute to improved energy accounting and optimisation strategies in HPC environments without requiring process isolation, marking a significant advancement in energy-aware computing. In future work, we aim to intensify the testing of our tool across a broader range of machines and experiments. We also plan to incorporate additional variables like frequency, fan speed, and GPU utilization into a deep learning model to enhance robustness and applicability in real-world scenarios. These improvements will contribute to developing more comprehensive and adaptive energy management solutions in high-performance computing.

\begin{credits}
\subsubsection{\ackname} Results achieved with the funding obtained under Axis IV of the PON Research and Innovation 2014-2020 "Education and research for recovery - REACT-EU". Experimental results correspond to the ORFEO Supercomputer at AREA Science Park. The project was partly funded by eXact Lab s.r.l.

\subsubsection{\discintname}
The authors have no competing interests to declare that are relevant to the content of this article.
\end{credits}
%
%
%
\bibliographystyle{splncs04}
\bibliography{references}
\end{document}